\documentclass[manuscript]{aastex}

\usepackage[usenames]{color}
\newcommand{\kms}{km\,s$^{-1}$}

\shorttitle{PDRs in NGC\,253}
\shortauthors{Mart\in et al.}

\begin{document}

%\title{Photodissociation traces in NGC253}
\title{Photodissociation chemistry footprints in the Starburst galaxy NGC\,253}

\author{Sergio Mart\'{\i}n}
\affil{Harvard-Smithsonian Center for Astrophysics, 60 Garden St.,  02138, Cambridge, MA, USA}
\email{smartin@cfa.harvard.edu}

\author{J. Mart\'in-Pintado}
%\affil{Departamento de Astrofis\'ica Molecular e Infrarroja, Instituto de Estructura de la Materia, CSIC, Serrano 121, E-28006 Madrid, Spain}
\affil{Centro de Astrobiolog\'ia (CSIC-INTA), Ctra de Torrej\'on a Ajalvir, km 4,
28850 Torrej\'on de Ardoz, Madrid, Spain}

\and

\author{S. Viti}
\affil{Physics and Astronomy Department, University College London, Gower Street, London, WC1E 6BT, UK}

\begin{abstract}
UV radiation from massive stars is thought to be the dominant
%Photodissociation is thought to be the main 
heating mechanism of the nuclear ISM in the late stages of evolution of starburst galaxies, creating large photodissociation regions (PDRs)
and driving a very specific chemistry.
We report the first detection of PDR molecular tracers, namely HOC$^+$, and CO$^+$, and confirm the detection of the also PDR tracer HCO
towards the starburst galaxy NGC\,253,
claimed to be mainly dominated by shock heating and in an earlier stage of evolution than M\,82, the prototypical extragalactic PDR.
%Though the three species are clearly detected, 
Our CO$^+$ detection suffers from significant blending to a group of transitions of $^{13}$CH$_3$OH,
tentatively detected for the first time in the extragalactic interstellar medium.
These species are efficiently formed in the highly UV irradiated outer layers of molecular clouds, as observed in the late stage nuclear starburst in M\,82.
The molecular abundance ratios we derive for these molecules are very similar to those found in M\,82.
This strongly supports the idea that these molecules are tracing the PDR component associated with the starburst in the nuclear region of NGC\,253.
%photodissociation
%is also a leading driver of the chemistry in NGC\,253.
The presence of large abundances of PDR molecules in the ISM of NGC\,253, which is dominated by shock chemistry, clearly illustrates the potential
of chemical complexity studies
% use of chemical tracers 
to establish the evolutionary state of starbursts in galaxies.
A comparison with the predictions of chemical models for PDRs shows that the observed molecular ratios are tracing the outer layers of UV illuminated
clouds up to two magnitudes of visual extinction.
%is presented.
We combine the column densities of PDR tracers reported in this paper with those of easily photodissociated species, such as HNCO, to derive the fraction
of material in the well shielded core relative to the UV pervaded envelopes.
Chemical models, which include grain formation and photodissociation of HNCO, support the scenario of a photo-dominated chemistry as an explanation to the
abundances of the observed species.
From this comparison we conclude that the molecular clouds in NGC\,253 are more massive and with larger column densities than those in M\,82, as expected
from the evolutionary stage of the starbursts in both galaxies.
\end{abstract}


\section{Introduction}
Intense UV radiation from massive stars
%Photodissociation 
is one of the main mechanisms responsible for the heating of the interestelar medium in the nuclear region of starburst galaxies.
This mechanism is particularly important in the latest stages of starburst (SB) galaxies where the newly formed massive star clusters are
responsible for
%the high UV radiation fluxes.
creating large photodissociation regions (PDRs).
This is the case for the prototypical SB galaxy M\,82, where the large observed abundances of molecular species such as HCO, HOC$^+$, CO$^+$, and
H$_3$O$^+$ are claimed to be probes of the high ionization rates in large PDRs formed as a consequence of its extended evolved nuclear starburst
\citep{Burillo02,Fuente06,VdTak08}.
%,Fuente05

Observational evidences point to a significant enhancement in the abundance of HOC$^+$ in regions with large ionization fractions.
The abundance ratio [HCO$^+$]/[HOC$^+$]$=270$ is found in the prototypical Galactic PDRs of the Orion Bar \citep{Apponi99}.
Similar or even lower abundance ratios are observed in the PDRs NGC\,7023 \citep[50-120,][]{Fuente03}, Sgr\,B2(OH) and NGC\,2024
\citep[360-900,][]{Ziurys95,Apponi97}, and the Horsehead \citep[75-200][]{Goico09}, as well as in diffuse clouds \citep[70-120,][]{Liszt04}.
This is in contrast with the much larger ratios of $\gg 1000$ found in dense molecular clouds well shielded from the UV radiation.
However, these low HCO$^+$/HOC$^+$ ratios are not found in other galactic PDRs. 
%contrary to this picture, 
Large values of this ratio of $\gtrsim2000$ are found in the PDRs M17-SW, S140, and NGC2023 \citep{Apponi99,Savage04}.
%as well as the low ratio of 460 in the ultracompact HII region in Mon R2 \citep{Rizzo03}.
The HCO molecule has also been observed to be a particularly good tracer of the PDR interfaces.
Low ratios of [HCO$^+$]/[HCO]$\sim2.5-30$ are found in prototypical Galactic PDRs \citep{Schene88,Schilke01}.
The large HCO abundance ($>10^{-10}$) altogether with the low ratio [HCO$^+$]/[HCO]$\sim1$ in the Horsehead PDR is claimed to be a
diagnostic for an ongoing FUV-dominated photochemistry \citep{Gerin09}.
CO$^+$ is also claimed to be particularly prominent in the chemical modeling of PDRs and high abundances of this molecule appear to be correlated
to similar enhancements of HOC$^+$ \citep{Sternberg95,Savage04}.
[CO$^+$]/[HOC$^+$] ratios in the range of 1-10 are observed in a number of PDRs \citep{Savage04}, but only of $\gtrsim 0.1.$ towards
the Horsehead PDR \citep{Goico09}.

As mentioned above, this set of PDR probes has been extensively studied towards M\,82.
However, no such complete studies have been carried out towards other prototypical galaxies, but for the detection of
HCO and HOC$^+$ towards NGC\,1068 \citep{Usero04} and H$_3$O$^+$ in Arp\,220 \citep{VdTak08}.
%Like M\,82, the starburst galaxy NGC\,253 is one of the brightest prototype of nearby starburst galaxies,
%with large mass of gas fueling the ongoing star formation at a rate of $2.8\,M_\odot \rm yr^{-1}$ \citep{Ott05,Minh07}.
M\,82 and NGC\,253 are the brightest prototypes of nearby SB galaxies, at a similar distance and showing very similar IR luminosities and
star formation rates of about $\sim3\,M_\odot \rm yr^{-1}$ \citep{Ott05,Minh07}.\
However, both galaxies show very different chemical composition. The chemistry and to a large extend the heating in the central region
of NGC\,253 is believed to be dominated by large scale low velocity shocks \citep{Martin06b}.
The similar chemical composition found in the nuclear region of NGC\,253 to that in Galactic star forming molecular complexes
points to an earlier evolutionary stage of the starburst in this galaxy than that in M\,82 \citep{Martin03,Martin05,Martin06b}.

Furthermore, our recent observations of the PDR component as traced by the easily photodissociated HNCO molecule towards a sample of galaxies
\citep{Martin08} showed the non-detection of HNCO in M\,82, at a very low abundance limit.
This low HNCO abundance supports the scenario that the PDR chemistry
dominates the molecular composition of the ISM in this galaxy.
However, from the HNCO measured abundance in NGC\,253, it would be placed in an intermediate stage of evolution
where photodissociation should be starting to play a significant role in driving a UV-dominated chemistry which has not been yet identified towards
this galaxy.
%than previously suspected for this galaxy.
The presence of a significant PDR component in NGC\,253 claimed from the HNCO abundance is also inferred from the similar intensity of the atomic
fine structure line intensities from PDR tracers like CII and OI \citep{Carral94,Lord96} observed in both M\,82 and NGC\,253.

In this paper we present the first detection of PDR molecular tracers HOC$^+$ and CO$^+$,
and confirm the detection HCO \citep[tentatively detected by][]{Sage95}
in the central region of NGC\,253 which allows the evaluation of the influence
of the photodissociation radiation in the nuclear ISM of this SB galaxy.
The results presented here support the scenario of the presence of a significant PDR component and clearly show the potential of molecular complexity
in estimating the
%to derive the
contribution of the different heating mechanisms of the ISM in the nuclei of galaxies.

\section{Observations and Results}
\label{sec.Obs}
The observations presented in this paper were carried out at the IRAM 30\,m and JCMT telescopes on Pico Veleta, Spain, and Mauna Kea, USA, respectively.

\subsection{IRAM 30\,m}
The IRAM 30\,m observations were performed in symmetrical wobbler switched mode with a frequency of 0.5\,Hz and a beam throw of
$4'$ in azimuth.
The $516\times1$\,MHz filter banks were used as spectrometers.

We have observed the transitions of $\rm C^{18}O$ $J=1-0$ (109.782\,GHz), HCO$^+$ $J=1-0$ (89.188\,GHz), HOC$^+$ $J=1-0$ (89.487\,GHz) and $3-2$ (268.451\,GHz), and the 
HCO $1_{0,1}-0_{0,0}$ (86.670\,GHz).
Beam sizes at these frequencies were $22''$, $28''$, and $9''$.
The nominal position for the observation was
$\alpha_{J2000}=00^{\rm h}47^{\rm m}33\fs3, \delta_{J2000}=-25^\circ17'23''$ for HCO$^+$ and HOC$^+$,
matching up the position used for the 2\,mm line survey \citep{Martin06b}.
The data on C$^{18}$O and HCO were centered
%extracted from a large raster map covering the central region of NGC\,253 which will be published elsewhere.
%In this paper we selected the closest spectrum, 
at  $\alpha_{J2000}=00^{\rm h}47^{\rm m}33\fs5, \delta_{J2000}=-25^\circ17'27''$.
The two positions are separated by $<5''$ which, considering the 3\,mm beam sizes, should have a negligible influence in the relative intensities.
Double Gaussian profiles have been fitted to all observed transitions and the corresponding derived fitting parameters are summarized in Table~\ref{tab:gaussfit}.

\begin{figure}
\includegraphics[angle=-90,width=\linewidth]{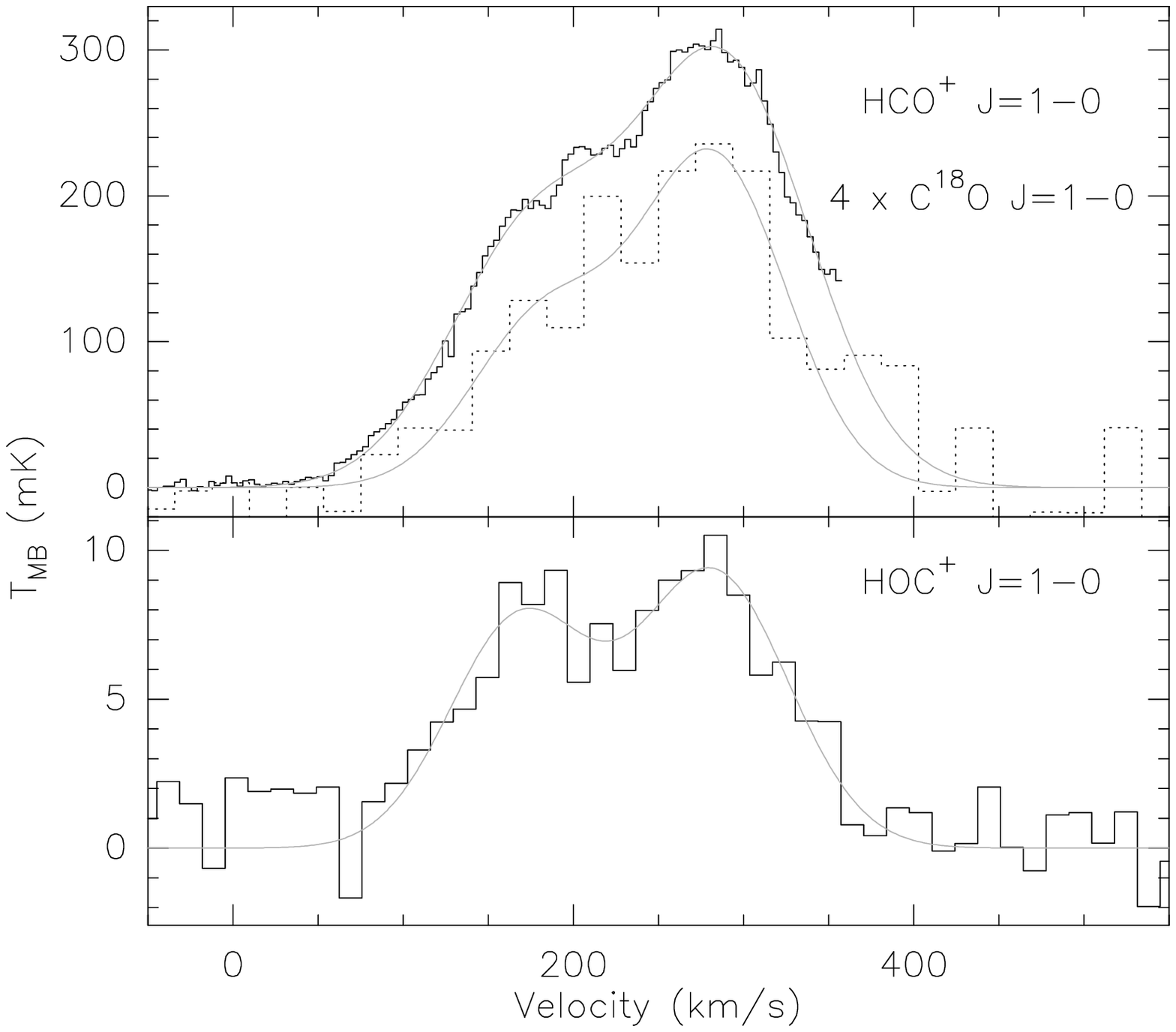}
\caption{IRAM 30\,m observations of the emission of HCO$^+$, $\rm C^{18}O$ and HOC$^+$ $J=1-0$ transitions with the fitted line profiles superimposed.
$\rm C^{18}O$ emission has been multiplied by a factor of 4 for comparison with the HCO$^+$ profile.
The spectral resolution are the original 1\,MHz ($\sim 3$ \kms) and smoothed to 4\,MHz ($\sim 13$ \kms), respectively. \label{fig:HCOp}}
\end{figure}

Fig.~\ref{fig:HCOp} shows the simultaneously observed $J=1-0$ features of HCO$^+$ and HOC$^+$
compared to the $\rm C^{18}O$ $J=1-0$ line profile.
Although at this frequency the SIS receivers image band rejection is larger than 20\,db, we observed the HOC$^+$ line tuned to two different velocities
(250 and 500 \kms) in order to confirm that the observed profile was not line emission coming from the upper side band.
Fig.~\ref{fig:HCOp} shows the average of both observations.
The HCO$^+$ $J=1-0$ was detected at the edge of the band covered in the 500 \kms\, tuning of HOC$^+$.
%The different shapes observed between the HCO$^+$ and HOC$^+$ is attributed to a small change in the pointing position.
Although pointing accuracy was of the order of $3''$, the different shapes observed between the HCO$^+$ and HOC$^+$ 
is attributed to a small change in the pointing position during the two observations.
%, as observed by a similar change in shape between the two observations of HOC$^+$.
This effect accounts for an uncertainty in the integrated intensity of $<10\%$.
Even though the HCO$^+$ feature was not completely covered by the backend, the integrated intensity we derive is in agreement within
$5\%$ with that of theobserved by \citet{Nguyen92} and we observe the line shape to be consistent with that of $\rm C^{18}O$.
As indicated in Table~\ref{tab:gaussfit}, only a coarse upper limit to the detection of HOC$^+\,J=2-1$ was obtained.

\begin{figure}
\includegraphics[angle=-90,width=\linewidth]{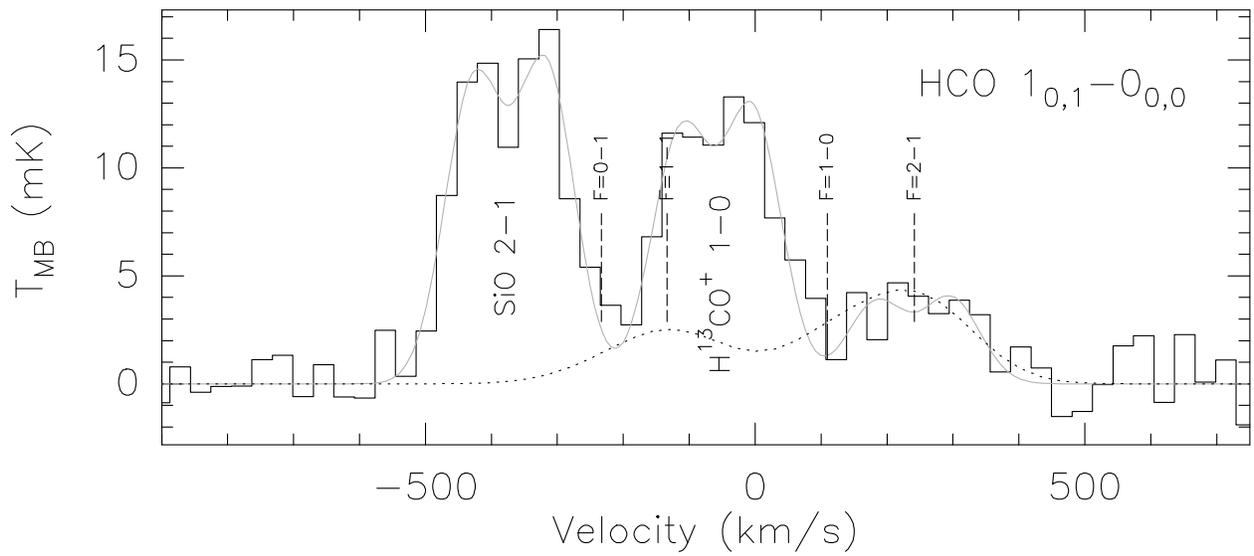}
\caption{HCO profile in the same window as SiO and H$^{13}$CO$^+$ observed with the IRAM 30\,m..
On top it is shown the triple Gaussian profile fitted to the transitions.
Dashed lines indicate the position of each hyperfine structure transition of HCO, where only the brighter one has been fitted.
The theoretical synthetic spectrum of HCO is shown with a dotted line (see Sect.~\ref{sec.Obs} for details).
Velocity resolution has been degraded to $\sim$31\,\kms.
\label{fig:HCO}}
\end{figure}

The HCO $J=1-0$ emission was observed in the same window as SiO $2-1$ and H$^{13}$CO$^+\,1-0$ and appears slightly blended to the latter.
With a significantly improved signal-to-noise ratio, we confirm the previous tentative detection of this HCO transition reported by \citet{Sage95}
with the NRAO 12\,m telescope.
Moreover, using the main beam brightness temperature from \citet{Sage95} of $\sim1$\,mK with at $72''$ beam and our
observed $\sim4$\,mK with a $28''$ beam we can make an estimate of the emitting source extent of $>20''$.
The double Gaussian profiles fitted to each species were constrained to have similar linewidths.
% forcing the linewidthsthe observed profiles assuming similar linewidths.
The resulting fitted line positions agree within the errors to those expected from the rest frequencies of each line.
Fig.~\ref{fig:HCO} shows the results of the fit superimposed on the observations
as well as the position of the hyperfine structure lines of HCO.
Only the brightest of the group ($F=2-1$) has been taken into account for the fit.
Assuming optically thin emission, the $F=1-0$ and $F=1-1$ transitions (at 86.708 and 86.777\,GHz)
are expected to show an intensity half of the main transition but they are completely blended to the
H$^{13}$CO$^+$ emission.
The $F=0-1$ transition at 86.805\,GHz is expected to be even fainter by a factor of 5, well below our detection limit.
Fig.~\ref{fig:HCO} shows in dotted line a synthetic spectrum of HCO assuming one velocity component centered at 255\kms\, with a linewidth
of 192\kms\,(as derived if only one component is fitted to the spectrum from the other lines)
and a peak intensity of the HCO $F=2-1$ line of 3.6\,mK.
% (10\% below the results derived from the Gaussian fits).
This shows that the fainter HCO hyperfine transitions may account for up to a $10-20\%$ of the H$^{13}$CO$^+$ integrated intensity.
%By not taking into account the emission from all the HCO hyperfine lines, the integrated intensity of H$^{13}$CO$^+$ might be overestimated on a $10-20\%$.

\subsection{JCMT}
JCMT observations were performed in beam switched mode with a frequency of 1\,Hz and beam throw of $2'$ in azimuth.
The ACSIS digital autocorrelator spectrometer was used with a bandwidth of 1600\,MHz providing a resolution of $\sim1$\,MHz.

We have used the receiver A3 to observe the CO$^+$ transition at 236.062\,GHz. At this frequency, the beam size of the telescope is $21''$
and the main beam efficiency 0.69.
The observations were carried towards the nominal position
$\alpha_{J2000}=00^{\rm h}47^{\rm m}33\fs1, \delta_{J2000}=-25^\circ17'18''$
\citep[radio continuum position,][]{Douglas96}.
As seen in the HC$_3$N $J=25-24$ profile, most of the emission is observed from one of the velocity components at this position,
which is due to the JCMT observed position being $\sim6''$ and $10''$ away from those observed with the IRAM 30\,m, respectively.
This position is half beam away from the positions observed with the IRAM\,30m, so the abundance ratios derived from this observation
might be affected by a larger uncertainty of up to a factor of 2.
However, this effect might be attenuated by the emission being extended over scales of $>20''$.

The CO$^+$ $5/2- 3/2\,F=2-1$ and $3/2- 1/2\,F=2-1$ transitions are clearly detected above the noise level ($\sim 1.5$\,mk in 30\,\kms channels).
However, we observe its profile significantly blended to that of the group of transitions of $\rm^{13}CH_3OH\, J=5-4$.
This overlap was not a problem in the case of the CO$^+$ detection towards M\,82 due to the significantly lower abundance of CH$_3$OH towards this
galaxy \citep{Martin06a}.
The CO$^+$ $3/2- 3/2\, F=2-1$ component was not detected due to its low relative intensity.
We have fitted the observed profile with a single Gaussian component CO$^+$ synthetic spectra with radial velocity and linewidth fixed from
those derived from HC$_3$N.
The relative intensities of the CO$^+$ components were also fixed to those expected from optically thin emission under under local thermodynamic equilibrium 
conditions.
The derived line profiles parameters are presented in Table~\ref{tab:gaussfit}.
Additionally, we simultaneously fitted a synthetic $^{13}$CH$_3$OH spectra, using the 18 $^{13}$CH$_3$OH transitions in the observed frequency range,
using the same constraints as for the CO$^+$ line fit.
The fit reproduces most of the observed features and shows that the emission from $^{13}$CH$_3$OH may explain the observed non Gaussian
CO$^+$ profiles.
Only the parameters derived for the three most intense components in the group are given in Table~\ref{tab:gaussfit}.
This is the first time that the $^{13}$C isotopologue of methanol is detected towards an extragalactic source.
Regarding the accuracy of the fitted parameters presented in Table~\ref{tab:gaussfit},
the integrated line intensities derived for CO$^+$ and $^{13}$CH$_3$OH
are likely underestimated by $\sim20\%$ due to the baseline determination.
In the next Section, we discuss the detection of $^{13}$CH$_3$OH in the context of the derived abundances with respect to those of the main
methanol isotopologue.

\section{Molecular abundances and ratios}

We have estimated the fractional abundances of the newly observed species in NGC\,253
assuming optically thin emission, LTE conditions, and similar spatial distribution for all species.
%Calculations were done for the parameters of excitation temperature $T_{\rm ex}=15\pm5$\,K and source size of $10''$.
%$10''\pm5''$.
Under these assumptions, we have calculated the column densities of H$^{13}$CO$^+$, HOC$^+$, HCO, and CO$^+$ for
an excitation temperature $T_{\rm ex}=15\pm5$\,K and an estimated source extent for each velocity component of $10''$.
The $T_{\rm ex}=15\pm5$\,K is assumed based on the average rotational temperatures derived from most of the species detected
towards NGC\,253 \citep{Martin06b}. Indeed the non detection of HOC$^+\,3-2$ implies low excitation temperatures of $T_{\rm ex}\sim10$\,K.
Both the excitation temperature and the emission extent have an important impact in the absolute derived column densities by up to a factor
of 2, however, the fractional abundances and abundance ratios are mostly independent of these assumptions.
We assume that the emission extent is similar for all observed species.
Table~\ref{tab:abunRatios} presents the column densities and fractional abundance ratios with respect to H$_2$ for all the species.
The total H$_2$ column density has been derived from the C$^{18}$O column density for each velocity component assuming an isotopic ratio
of $\rm^{16}O/^{18}O=150$ \citep{Harrison99} and a CO/H$_2$=$10^{-4}$.
The relative abundances derived for all molecules are presented in Table~\ref{tab:abunRatios}.
Additionally, abundance ratio of H$^{13}$CO$^+$ with respect all species is also presented in Table~\ref{tab:abunRatios}.
The errors in the derived column densities take into account the statistical error of integrated intensities and the uncertainty in
the assumed excitation temperature. These errors are subsequently propagated to the abundance ratios.
%, that may also vary for each source
%adding additional uncertainty to the ratios.
%observed transition, assumed an isotopic ratio $\rm^{12}C/^{13}C=40$ \citep{Henkel93}.
The HCO$^{+}$ abundance has been derived from that of H$^{13}$CO$^{+}$ to avoid the opacity effects affecting the main isotopologue.
Indeed, if the $^{12}$C/$^{13}$C ratio of $\sim40$ derived for NGC\,253 and NGC\,4945 \citep{Henkel93,Henkel94,Martin05} applies to all these galaxies,
an average opacity of $\tau_{HCO^+\,J=1-0}\gtrsim1$ is derived from the $\rm HCO^+/H^{13}CO^+ \,J=1-0$ line ratio \citep[this work and][]{Usero04}.
As a consequence, ratios may be underestimated by a factor of $\sim 2$ if derived from the main isotopologue.
%the abundance ratios can are proportional to the line integrated intensity ratios for the observed transitions as
%$\rm [H^{13}CO^+]/[HCO]\simeq(0.059\pm0.007)\,I_{H^{13}CO^+\,J=1-0}/I_{HCO\,J=1-0}$
%and 
%$\rm [H^{13}CO^+]/[HOC^+]\simeq(1.62\pm0.01)\,I_{H^{13}CO^+\,J=1-0}/I_{HOC^+\,J=1-0}$.
%Abundances and ratios derived from our observations are shown in Table~\ref{tab:abunRatios}.
The CO$^+$ column density presents a significant uncertainty due to its blending with the newly detected isotopologue of methanol, $^{13}$CH$_3$OH.
However, the CO$^+$ abundance should not be affected by more than a factor of 2.

We have derived a column density for $^{13}$CH$_3$OH of $\sim1.6(1.0)\times10^{14}\,\rm cm^{-2}$.
If we compare this column density with that of the main isotopologue from \citep{Martin06b} it results in a  CH$_3$OH/$^{13}$CH$_3$OH ratio of $12\pm7$, significantly lower than the
ratio $^{12}$C/$^{13}$C=40 \citep{Henkel93}.
Both the difference of $6''$ in the observed positions and a different filling factor of CH$_3$OH
might account for part of this difference.
However, the integrated intensities measured for the methanol group of transitions at 145.1\,GHz by \citet{Martin06b} and \citet{Hutte97}
at positions differing by $>13''$ show a variation of $<6\%$ so the difference in positions is not likely to contribute to
this difference.
%That of different emitting regions covered by the different telescope beams might be the most plausible
%explanation for such difference.
Thus, opacity is likely the dominant effect as observed in the Galactic center \citep[GC,][]{Requena06}.
From the $^{12}$C/$^{13}$C ratio in NGC\,253 we derive a fractional abundance of methanol of $\sim$10$^{-7}$, close to the abundances observed in the 
Galactic center \citep{Requena06}.
Methanol is, after CO and NO, the most abundant molecule in the nucleus of NGC253 \citep{Martin06b}.
% (1950) 0:45:06.0	-25:33:36	Hüttemeister, S., Mauersberger, R., Henkel, C. 1997, A&A, 326, 59
% (2000) 00h47m33.38022s   -25d17m13.9945s
% Difference:
% R.A 0.28s=4.2''->3.8''   dec=4'' =>5.5
%FREQUENCY	Tmb dv	           Tmb	               Vlsr	Av1/2
%145.1	21.4 (1.8)	102 (25)	               213 (8)		196 (21)
% 2mm SURVEY
% Line      Area               Position           Width              Intensity
%  1   20221.     (210.832)  245.330 (  1.046)  197.500 (  2.222)   96.183

\section{Discussion: The PDR component in NGC\,253}
\subsection{NGC\,253 in context}
\label{sec.rat}

Table~\ref{tab:ratios} shows the HCO$^+$/HOC$^+$, HCO$^+$/HCO, and  HCO$^+$/CO$^+$ abundance ratios resulting from our
measurements in NGC\,253 compared to those of the similar SB galaxy M\,82, and the Seyfert 2 with nuclear SBs, NGC\,1068, and NGC\,4945,
together with prototypical galactic PDRs, where observations of these species have been reported.
For the shake of consistency, the abundance ratios of the other galaxies have been calculated from the available line profiles
obtained from the literature.
As already explained before, for this comparison we used H$^{13}$CO$^+$ to derive the HCO$^+$ column densities in order to avoid the opacity effects.
We have assumed the isotopic ratio of $^{12}$C/$^{13}$C=40 \citep{Henkel93} to derive the column densities of the main isotopologue.
For M\,82, the $\rm HCO^+/HOC^+$ and $\rm HCO^+/CO^+$ abundance ratio have been derived from the observations
by \citet{Fuente06} towards the Eastern molecular lobe.
Our measured ratios are
%with respect to HCO$^+$ are 
$\sim59$ and $\sim0.8$, which are $>30\%$ larger that the ratios derived by \citet{Fuente06}.
%the single dish data of H$^{13}$CO$^+$ \citep{Mauers91} and HOC$^+$
%\citep[Western position by][]{Fuente05}.
%The two observations were pointed $<7''$ away from each other, which is a relatively small error within the 28$''$ telescope beam.
% Fuente 09:55:51.9 69:40:47.11         -14'',-5''  Western
% Rainer 09:55:49.7 69:40:36.2   (-2.2s)-11.3'',-10.9''   -> differencia 3.7'',4.1''=5.5''
%Assuming the $^{12}$C/$^{13}$C$=40$ ratio, the values in Table~\ref{tab:ratios} can be scaled to the ratios with HCO$^+$.
%The resulting $\rm HCO^+/HOC^+$ ratio of $\sim75$ is a $50\%$ larger than the ratio
%more than a factor of 2 higher
% 2.5 exactamente
%derived by \citet{Fuente05}.
This difference is due to the significant missing flux of the H$^{13}$CO$^+$ interferometric maps \citep{Burillo02} they used for comparison, as well
the higher $^{12}$C/$^{13}$C ratio they used to compare with single dish HCO$^+$ observations.
By comparing the convolved integrated intensity from the H$^{13}$CO$^+$ interferometric maps by \citep{Burillo02}
towards the western lobe \citet{Fuente06} and the single dish data of by \citet{Mauers91} towards a nearby position,
we have estimated a $\gtrsim50\%$ missing flux.
Thus we used the H$^{13}$CO$^+$ $1-0$ data by \citet{Mauers91} in our measured ratios.
%by \citet{Fuente06} towards the western molecular lobe and 
%, even with the higher $^{12}$C/$^{13}$C ratio they used for their calculations, 
%even with the higher $^{12}$C/$^{13}$C ratio they used {\bf ?!?!}.
The ratios towards NGC\,1068 were derived for the regions within the galaxy where lines intensities were tabulated by \citet{Usero04}.
%However, the $\rm HCO^+/HOC^+$ abundance ratio derived under LTE conditions is a factor of 3 larger than that derived by \citep{Usero04}
%using the LVG approximation {\bf ?!?!}.
%3.2 exactamente
The observations from \citet{Wang04} were used to derive the $\rm HCO^+/HCO$ ratio towards NGC\,4945.

\subsubsection{PDR abundance ratios in SB galaxies}
We find that the three derived HCO$^+$/HOC$^+$, HCO$^+$/HCO, and HCO$^+$/CO$^+$ abundance ratios in the two starbursts, NGC\,253 and M\,82,
are equivalent within the measurement errors.
The ratios are also in reasonable good agreement to those found in galactic sources with similar FUV fluxes (see Table~\ref{tab:ratios}).
Such high abundances ratios of HOC$^+$, HCO, and CO$^+$ relative to HCO$^+$ have been claimed to be the evidence of M\,82 being mostly
dominated by photodissociation.
%If that is the case, these comparisons 
We notice that the average $\rm HCO^+/HCO$ 
%and $\rm H^{13}CO^+/CO^+$ 
ratio in NGC\,253 are even lower than that measured in M\,82.
In the case of HCO, the interferometric maps of M\,82 clearly resolve the spatial variations in this ratio across the galaxy nuclear region.
However, 
%if we compare the PDR dominated regions in M\,82 where we find the lowest 
towards the region of peak HCO emission in the M\,82 maps, we find a ratio of $\rm H^{13}CO^+/HCO\sim0.12\pm0.04$, 
%this value is still 
equivalent to the average observed towards NGC\,253.
Our data show that the ISM in the nuclear region in NGC\,253 must be significantly pervaded by a strong UV radiation flux from the 
massive star clusters formed in the starburst, as also suggested by the study of the abundances of the HNCO/CS ratio \citep{Martin09}.
Moreover, these new observations would imply that photodissociation plays a similar role in the ISM heating of both NGC\,253 and M\,82.

\subsubsection{PDR abundance ratios in AGN galaxies}
\label{sect.PDR_AGN}
Both $\rm HCO^+/HCO$ and $\rm HCO^+/HOC^+$ abundance ratios in NGC\,1068 are different by a factor $2-3$ from those of NGC\,253 and M\,82.
Furthermore, $\rm HCO^+/HCO$ is also found to be up to a factor of $\sim2$ lower in the ring of star formation
than towards the nuclear region.
%, and is equal to that found in the PDRs in M\,82.
Like in NGC\,253, these ratios are consistent with the decrease in the abundance of molecules such as HNCO from the nuclear region
to the starburst ring \citep{Martin09}.
The tentative detection of HCO in the circunnuclear disk (CND) of NGC\,1068
%\citep[see ($0'',0''$) position in Fig.~1 from][]{Usero04}, 
\citep{Usero04}, 
suggest a rough $\rm H^{13}CO^+$ to HCO line intensity ratio of $\sim 2-3$, which turns into an abundance ratio in the range of $\sim 0.09-0.13$,
closer to the values derived in SB dominated galaxies.
This value at the CND can be significantly biased by the emission from the star forming ring covered at half-power by the $28''$ beam.
Therefore, it is clear that this ratio does not significantly decrease towards the nuclear AGN in this galaxy with the typical angular
resolution of $20''-30''$.
On the other hand, the ratio $\rm HCO^+/HOC^+$ is only a factor of 2 lower in the starbursts galaxies than towards the nuclear AGN in NGC\,1068.
From these observations, it is unclear whether photodissociation does play a major role in the AGN dominated center of NGC\,1068.
Unfortunately, no observations of HOC$^+$ are available towards the SF ring in this galaxy.

Similar to NGC\,1068, the obscured Seyfert 2 nucleus in NGC\,4945 is surrounded by a starburst ring more prominent than in NGC\,1068
\citep{Genzel98}.
The HCO$^+$/HCO ratio found in NGC\,4945 is even lower than that found in the other galaxies.
However, the fit to these lines was claimed to be very uncertain by \citet{Wang04}.

\subsection{Comparison to PDR chemical models}
\label{Sec.Models}

We have compared our observed fractional abundances and abundance ratios with those predicted by the UCL\_PDR model \citep{Bell06}.
%(Bell et al. 2006; Bell 2007).
The UCL\_PDR code is a time and depth dependent one dimensional PDR model that simultaneously solve the chemistry, thermal balance and radiative transfer
within a cloud \citep[see][for more details]{Bell06}.
We have adopted a hydrogen density of $\rm 10^5\,cm^{-3}$, a radiation field $G_0\sim5000$ in units of Habing field,
and a cosmic radiation rate of $10^{-16}$.
The high density of $\rm 10^5\,cm^{-3}$ is derived from the multiline analysis of CS and HC$_3$N \citep[][; Aladro et al. In Prep]{Bayet08,Bayet09}.
Our estimates of the cloud structure (see Sect.~\ref{sec.GMCinSB}) depends on the averaged radiation field which might be different in both galaxies.
The averaged radiation fields in both NGC\,253 and M\,82 have been inferred from the fine structure lines
and they are of $2\times10^4$ and $10^3$, respectively, with large errors of a factor of 2 \citep{Carral94,Lord96}.
This would imply that the PDR envelope should be larger in the clouds of NGC\,253 than in M\,82.
Since we do not aim to quantitatively model the particular abundances measured in NGC\,253 but to investigate the physical 
conditions that would give rise to the wealth of observed molecules in starburst, the
value of $G_0=5\times10^3$ used is a geometric mean value derived from the fine structure lines in both galaxies.
Given that our estimates of the $A_{\rm v}$ for the PDR are based on this geometric mean, the expected changes in
$A_{\rm v}$ for the two galaxies would be just a factor of 1.6.
%The value of $G_0$ used is a mean value derived from the fine structure lines in M\,82 and NGC\,253 \citep{Carral94,Lord96}.

\begin{figure}
\includegraphics[angle=-90,width=\linewidth]{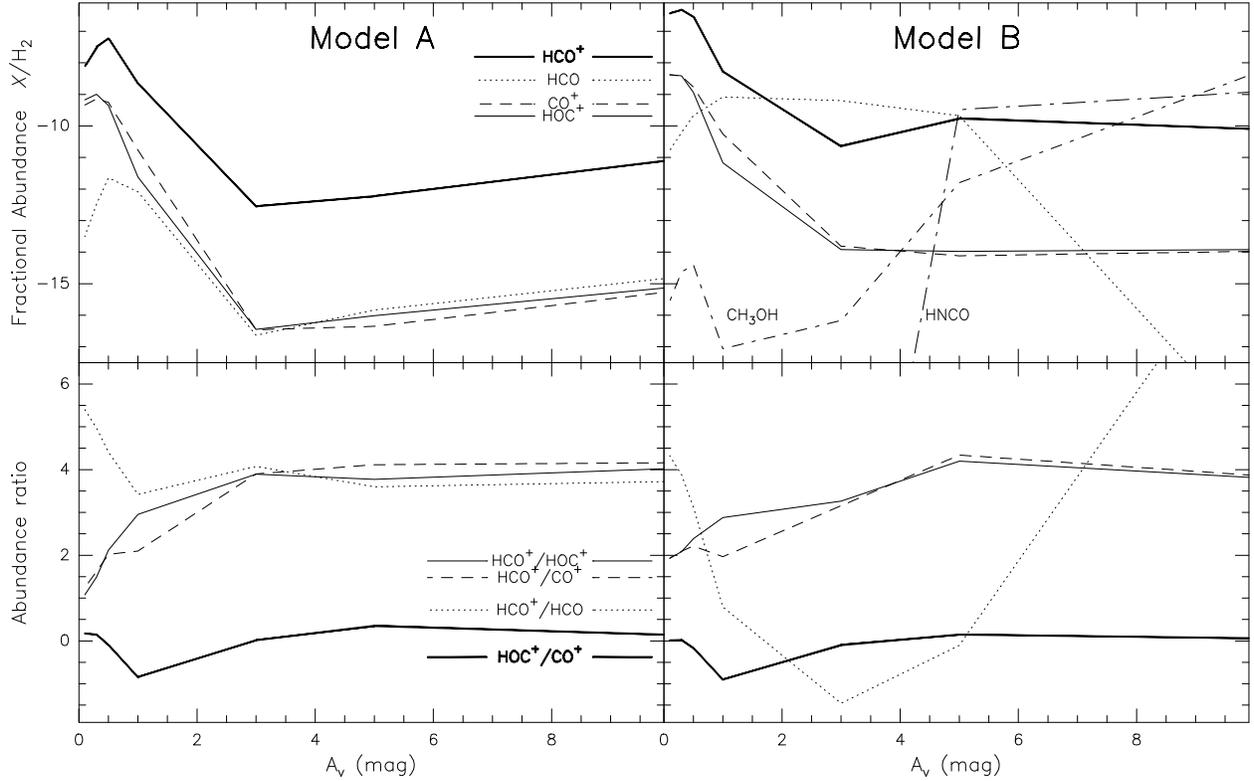}
\caption{
Theoretical predictions for the fractional abundances relative to H$_2$ ({\it Upper panels}) and abundance ratios ({\it Lower Panels})
for the observed species as derived from the two different PDR models:
a pure gas-phase model A ({\it Left panels}) and a coupled dense core-PDR model B ({\it Right panels}).
%Abundances, column densities and ratios for the observe species as derived from the PDR chemical models described in the text.
%Continuous and dashed lines represent the results from the pure gas-phase PDR model (A) and coupled dense core-PDR model (B).
Details are given in Section~\ref{Sec.Models}.
The vertical position of the key for each molecule and ratio shown only in the plots for model A correspond to the actual derived parameters
from the observations.
Additionally, the fractional abundances of CH$_3$OH and HNCO are shown for model B.
% are shown as horizontal lines.
\label{fig:PDRmod}}
\end{figure}

We have ran two different models:
Model A is a standard time dependent gas-phase PDR model where the initial composition is atomic;
while Model B is computed using a coupled dense core-PDR model where the diffuse material, initially also purely atomic and gaseous, collapses to reach
a final density of $\rm 10^5\,cm^{-3}$. During the collapse the gas depletes on the grains forming icy mantles which remain on the dust until irradiation
from a UV field is switched on, evaporation occurs and the typical PDR chemistry takes place.
In both models the temperature is calculated self-consistently at each depth and time step by thermal balance.
Fig.~\ref{fig:PDRmod} shows the predicted abundances and abundance ratios as a function of visual extinction ($A_{\rm v}$) for Model A (left panels)
and B (right panels).
HNCO and CH$_3$OH results are only shown for Model B in Fig.~\ref{fig:PDRmod}.

Observed abundances of HCO$^+$ and HOC$^+$ towards NGC\,253 (shown as horizontal lines in the key of Fig.~\ref{fig:PDRmod}) 
are well reproduced by the models for very low extinction of $A_{\rm v}\sim1-2$.
The HCO abundance observed is a factor $2-6$ above the maximum predicted by the model B.
It is important to take into account that while we assumed a ratio CO/H$_2$=10$^{-4}$ to calculate the fractional abundances, this ratio is not
constant in the models and hardly ever reaches this value.
%However, this is well within the uncertainties of the column density of molecular hydrogen and therefore the absolute fractional abundances derived.
On the other hand, the abundance ratios of the observed molecules, unaffected by the hydrogen determination uncertainty,
agree well with the model predictions.
The model shows that the abundances of CO$^+$, follows the same pattern as HOC$^+$.
The correlation between these two molecules was also predicted by previous theoretical studies \citep{Sternberg95,Savage04}.
Thus, CO$^+$ measurement allows us to confirm the effect of photodissociation suggested by the large abundance of HOC$^+$.

The ions HOC$^+$, CO$^+$ and HOC$^+$ are mostly formed at the edge of the cloud and while the two models predict similar abundances for these ions,
it is worth noting that the observed abundances of other species such as HNCO, CH$_3$OH, or HOCO$^+$ can only be explained with Model B.
However, HNCO and CH$_3$OH only reach observable abundances for $A_{\rm v}\sim5$ magnitudes.
This implies that, while photodissociation does play an important role in the chemistry of NGC253, the molecular clouds affected by the UV radiation must
contain a dense core well shielded from the UV radiation and rich with gas phase icy mantle molecules like HNCO and CH$_3$OH.

%Fig.~\ref{fig:PDRmod} shows the predicted abundances and abundance ratios as a function of the visual extinction ($A_{\rm v}$).
%The species HCO$^+$, CO$^+$, and HOC$^+$ are mostly formed in the edge of the molecular cloud, for $A_{\rm v}<2$, while HCO is most efficiently 
%formed deeper into the cloud in the region $1<A_{\rm v}<5$.
%
%For the modeling we have adopted 
%For these molecules, we show the result of two different physical models: 
%(a)pure time dependent PDR models where the initial composition was purely gaseous, shown in a continuous line, 
%and (b) a 2-phase model where the dense gas forms from diffuse, purely atomic gas, freezes out until it reaches the desired density and
%then the 'excitation' is 'switched on' in the sense that the model follows the chemical evolution of the gas 
%plus evaporated mantles with the temperatures and radiation field and cosmic ray ionization rate as listed in the tables {\bf put into figure}.

\subsection{The molecular clouds in SB galaxies}
\label{sec.GMCinSB}
\citet{Martin08} have used a comparison with the GC molecular clouds to propose another PDR diagnostic based on the relative abundance of HNCO to CS.
The large variation of the HNCO/CS abundance ratio between UV radiated clouds to those well shielded clouds only affected by shocks was interpreted
as the fast photodissociation of the fragile molecule HNCO, efficiently produced on the icy mantles and delivered into gas phase by low velocity shocks.
Similarly, in a sample of nearby galaxies \citet{Martin09} found changes of nearly two orders of magnitude from the shock dominated chemistry in
M\,83 and IC\,342 to UV dominated chemistry in M\,82.
The extremely low HNCO abundance in M\,82 and the large abundances of HCO, HOC$^+$ and CO$^+$ support the idea that the HNCO/CS ratio is a measure
of the relative importance of the  UV heating to  shock heating and the evolutionary state of the starburst in galaxies.

The detection of HCO, HOC$^+$ and CO$^+$ in NGC253 with similar column densities and abundance to those in M82 suggest that the PDR component is similar
in both galaxies as suggested by the similar atomic fine structure line intensities in both galaxies.
The results of model B confirm the observational trends observed in HNCO and the other PDR molecules in galaxies.
For molecular clouds with $A_{\rm v}\sim1-2$ (i.e column densities of $1-2\times 10^{21}\rm cm^{-2}$) illuminated by a strong UV radiation field like
in the galaxies in our sample, 
HNCO and CH$_3$OH are largely photodissociated and only HCO, HOC$^+$ and CO$^+$ should be observed like in M82 \citep{Martin06a,Martin09}.
Then, not very massive molecular clouds and widely translucent to the UV radiation should dominate in M\,82.
On the other hand, for galaxies with massive molecular clouds (large visual extinction) or low UV radiation fields,
HNCO and CH$_3$OH are well shielded and the abundance ratio of HNCO/CS will reach its maximum value.
Considering that M\,83 and IC\,342 represent the stage of galaxies with an extremely low PDR component, the lower HNCO/CS ratios measured 
for NGC\,253, NGC\,4945 and NGC\,1068 indicate that the PDR component must be substantial, as observed in other PDR tracers. 

For NGC\,253 we find that the PDR component is similar to that in M82, but the total column density of dense gas is a factor of 
2-3 larger in NGC\,253 than in M\,82 from the low HNCO and CH$_3$OH abundance in the latter.
Considering the PDR column densities in both galaxies are similar, this component should represent about 
$1/3-1/2$ of the total molecular column density in NGC253.  This is roughly consistent with the decrease by a factor of $2-3$ of the HNCO/CS 
ratio as compared with that of M\,83 or IC\,342 \citep{Martin09}.
This suggests that the molecular clouds properties in M\,82 and NGC\,253 must be quite different in terms of the total molecular column density,
which implies that the sizes or the densities are different, or a combination of both.  Using the atomic fine structure and CO emission lines,
\citet{Carral94} and \citet{Lord96} have also proposed that the clouds in M\,82 and NGC\,253 are quite different.
The clouds in NGC\,253 are slightly smaller than those in M\,82, but with masses a factor of 15 larger than for M\,82.
The NGC\,253 average cloud column densities are therefore factor of 20 larger than in M\,82.
Though the total column densities of the M82 clouds inferred from the atomic fine structure lines
are a factor of 5 larger than those predicted from the PDRs tracers and
the HNCO abundance in this galaxy, similar constrains are derived both the molecular and the atomic tracers.
Therefore, the clouds in NGC\,253 are more massive than in M\,82.

We can even make a very rough estimate of the average properties and structure of the 
molecular clouds in NGC\,253 and M\,82 by combining the complementary 
information obtained from the PDR tracers presented in this paper and 
the HNCO column densities from \citep{Martin09}.
While HCO, HOC$^+$ and CO$^+$ mainly trace the PDR region up to $A_{\rm v}= 4-5$ (i.e. H$_2$ column 
densities of $5\times10^{21}\rm cm^{-2}$), the HNCO emission only arises from the well 
shielded core ($A_{\rm v}>7$) of the molecular clouds. The HCO, HOC$^+$ and CO$^+$ 
column densities in NGC\,253 and M\,82 indicate similar averaged column 
densities in the PDR envelopes of the molecular clouds in both galaxies. 
The big difference in the molecular cloud structure in both galaxies is 
in the size (H$_2$ column density) of the well shielded cores of the 
molecular clouds. In the case of M\,82, where HNCO has not been detected, 
we can set an upper limit to the HNCO column density of $7\times10^{12}\rm cm^{-2}$. 
This translates to a upper limit to the core H$_2$ column density of  $<3\times10^{20}\rm cm^{-2}$
for the HNCO fractional abundance of $2\times10^{-8}$ derived for the 
well shielded clouds in the galactic center \citep{Martin08}.
The averaged shielded cloud cores in M\,82 are smaller by more than one order
of magnitude than the PDR envelope. In the case of NGC\,253, the H$_2$ column
densities of the shielded cloud cores is $10^{22}\rm cm^{-2}$, a factor of 2 
larger than PDR envelope. Assuming a similar averaged density 
distribution in the molecular clouds in both galaxies, the clouds in 
NGC\,253 would be a factor $2-3$ larger than in M82.

\subsection{The contribution X-ray induced chemistry}
The PDR model presented by \citet{Fuente06} failed to reproduce the large CO$^+$ column density of a few $10^{13}\rm cm^{-3}$ observed towards M\,82.
This lead \citet{Spaans07} to explore the possibility of an enhanced X-ray induced chemistry in this galaxy.
\citet{Spaans07} concluded that such high formation of CO$^+$ can only be explained by X-ray irradiated molecular gas with densities of $10^3-10^5\rm cm^{-3}$.
Although the X-ray luminosity of NGC\,253 is a factor of $2-4$ below that of M\,82, both galaxies have a significant X-ray emission in the range
of $\sim10^{40}\rm\, erg\,s^{-1}$ \citep{Cappi99}.
Similar to M\,82, the NGC\,253 total CO$^+$ column density is $(3.6\pm1.1)\times10^{13}\rm\,cm^{-2}$.
Moreover, both show a similar HCO$^+$/CO$^+$ ratio of $\sim 30-40$.

The models presented in this paper are able to produce such column densities for visual extinctions of $A_{\rm v}\sim3-5$ for model A, and $A_{\rm v}\sim0.5-1$
for model B.
These models have been calculated with a radiation field, a cosmic ray flux and a H$_2$ density smaller by a factor of 2, 40 and 4, respectively,
with respect to those assumed in the models of \citet{Fuente06}.
Furthermore, we are able to reproduce the abundances and abundance ratios measured for all the other observed species presented in this paper.
\citet{VdTak08} showed how the measured abundance of $\rm H_3O^+$ in M\,82 can be both produced by PDR with a high cosmic-ray ionization
or by an XDR.
%Indeed, PDR models can draw similar results to X-ray models when an enhanced cosmic-ray ionization is used which, for most species, is a good
% enough approximation of the X-ray chemistry.
% However, as mentioned above, our models do not use particularly high cosmic ray fluxes.
Indeed, an increase of cosmic ray ionization rate in PDR models may be qualitatively used to simulate XDR-like environments.
Our models, however, do not use particularly high cosmic ray fluxes (a factor of 5 higher than standard).
Thus, though the X-ray irradiation is substantial in SB galaxies, the PDR models presented in this paper can reproduce the molecular abundances
observed towards the brightest prototypes, M\,82 and NGC\,253.
Moreover, no significant changes in the abundances of HOC$^+$ and HCO are found towards the nuclear AGN of NGC\,1068 where X-ray radiation is significantly
more important than in SB nuclei, as shown in Sect.~\ref{sect.PDR_AGN}.
Unfortunately, no CO$^+$ observation has been reported towards this Seyfert 2 nucleus.

\section{Conclusions: The pervading UV field in evolved starburst}

The comparison of model predictions with the observations presented show
%As the presented models have shown, 
that the abundance of the species observed in this work towards NGC\,253, namely HCO$^+$, CO$^+$, and HCO, are most efficiently formed in the outer region
of the molecular clouds where the gas is highly irradiated by the incident UV photons from massive stars.
The high molecular abundances derived for these species in NGC\,253 
%seems to rule out 
suggest that the PDR component in this galaxy is  similar to that found in
M\,82,
% as 
claimed to be the prototype of extragalactic PDR.
%Surprisingly 
The abundance ratios found for this limited sample of galaxies are of the same order as those observed towards galactic PDRs, which stress the
importance of photo-dominated chemistry in galaxy nuclei.
Large amounts of molecular material are affected by photodissociation not only in NGC\,253, but also towards
the star forming regions around the Seyfert 2 nuclei in NGC\,4945 and
%and the star forming ring around
%the nucleus of 
NGC\,1068.
This is consistent with the HNCO/CS ratio in these galaxies which suggest that a fraction of HNCO has been photodissociated in PDRs.
The combination of the observations of HCO, HOC$^+$ and CO$^+$ with that of HNCO seems to confirm that
their abundances reflect
%Whether this is purely an 
the evolutionary stage of the starbursts in these galaxies.
% or whether high degrees of photodissociation is a common
%symptom for starburst and more general for galactic nuclei, will be determined by extensive search of these molecular tracers over a wider galaxy sample.
Although photodissociation is the most likely scenario for the enhancement of the observed reactive ion in starburst environments, X-ray dominated
chemistry has been claimed to be responsible for the high abundances observed around AGNs in circunnuclear disk of NGC\,1068 \citep[HOC$^+$][]{Usero04} and
towards the ultra luminous infrared galaxy Arp\,220 \citep[$\rm H_3O^+$][]{VdTak08}.

Therefore, M\,82 is still outstanding not only as a PDR dominated galaxy, but by the
%long known 
underabundance of complex molecules such as CH$_3$OH, HNCO or SiO \citep{Mauers93,Martin06a,Martin06b},
%This low abundances have been claimed to be the 
evidence for the lack of large amounts of dense molecular material which would potentially
fuel its nuclear starburst as compared to other starburst galaxies like NGC\,253 \citep{Martin09}.
Our data in combination with the HNCO abundances \citep{Martin09} indicate that the molecular clouds
in M\,82 are different from those in NGC253.
Although having a similar overall PDR component, the clouds in NGC\,253 have to be more massive and have larger column densities those in M\,82.

\acknowledgments

 This work has been partially supported by the Spanish Ministerio de Ciencia e
 Innovaci\'on under project ESP2007-65812-C02-01, and by the ``Comunidad de Madrid'' Government under PRICIT project
 S-0505/ESP-0237 (ASTROCAM).

{\it Facilities:} \facility{IRAM 30m,JCMT}.

\clearpage

\clearpage

\clearpage

\begin{figure}
\includegraphics[angle=-90,width=\linewidth]{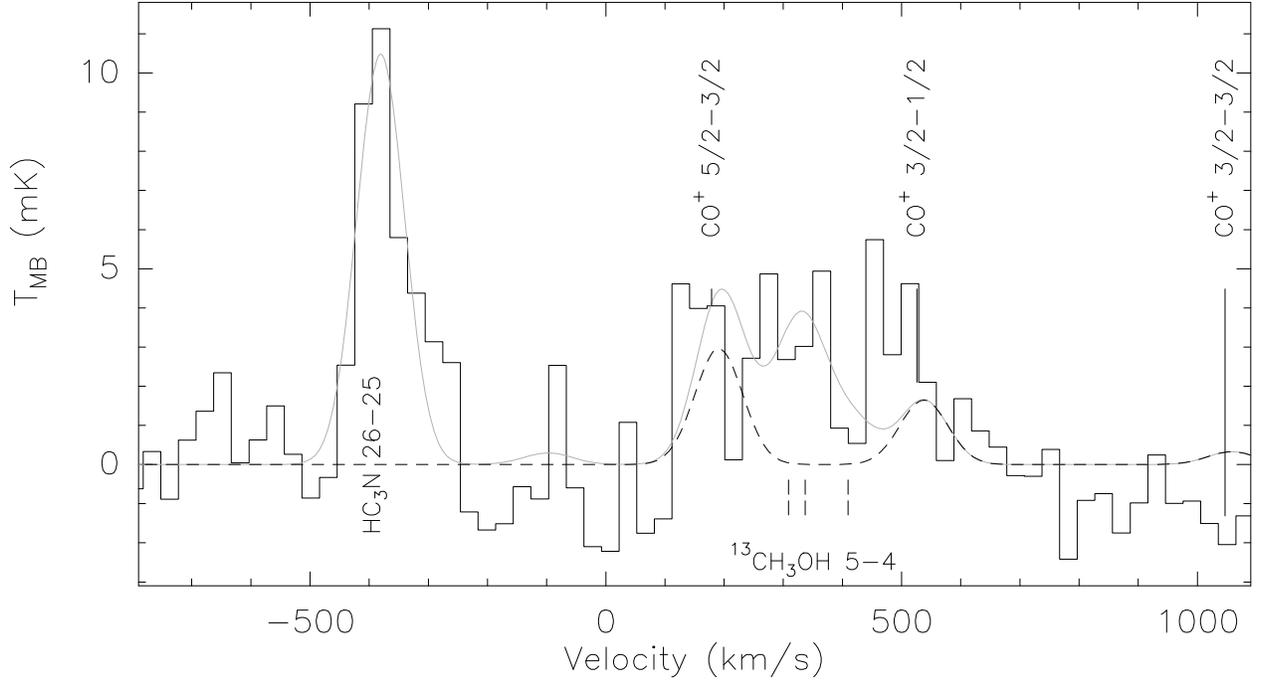}
\caption{JCMT observations of the CO$^+$ emission blended with transitions of $\rm^{13}CH_3OH\,J=5-4$ and observed in the same window as HC$_3$N\,$J=25-24$.
This is the first detection of the $^{13}$C isotopologue of methanol.
The overall spectral fitting to all the lines is shown in grey. The contribution of the CO$^+$ emission is shown with dashed line.
The position of the three brighter transitions of the $J=5-4$ group of $\rm^{13}CH_3OH$ are shown indicated with vertical dashed lines.
See Sect.~\ref{sec.Obs} for details on the fitting to the spectra.
Velocity resolution has been degraded to $\sim$30\,\kms.
\label{fig:COp}}
\end{figure}

\clearpage

\clearpage

\begin{table}
\begin{center}
\caption{Parameters derived from the observed line profiles. \label{tab:gaussfit}}
\begin{tabular}{l c c c c}
\tableline
\tableline
Transition             &  $\int{T_{\rm MB}{\rm d}v}$  & $v_{\rm LSR}$   &  $\Delta v_{1/2}$             & $T_{\rm MB}$ \\
                       &  (K\,km\,s$^{-1}$)           & (km\,s$^{-1}$)  &  (km\,s$^{-1}$)               & (mK)         \\
\tableline
$\rm C^{18}O\,1-0$     &  $3.1\pm0.6$                 &  $183\pm13$     &  $100\pm9$                    & 116.75      \\
                       &  $6.0\pm0.8$                 &  $283\pm7$      &  $100\pm9$\tablenotemark{a}   & 223.48      \\
HCO$^+\,1-0$           &  $21.98\pm1.1$               &  $177.9\pm0.4$  &  $118.1\pm0.3$                &  174.9       \\
                       &  $35.79\pm1.4$               &  $289.0\pm0.3$  &  $118.1\pm0.3$\tablenotemark{a}  &  284.8       \\
HOC$^+\,1-0$           &  $0.8\pm0.2$                 &  $170\pm10$     &  $100\pm20$                   &    7.6       \\
                       &  $1.0\pm0.2$                 &  $282\pm9$      &  $100\pm20$                   &    9.2       \\
HOC$^+\,3-2$           &  $<0.8$ \tablenotemark{b}    &                 &                               &  $<4.1$      \\
%HCO\,$1_{0,1}-0_{0,0}$ &  $0.87\pm0.12$               &  $250\pm20$     &  $192$\tablenotemark{a}       &    4.3       \\
HCO\,$1_{0,1}-0_{0,0}$ &  $0.41\pm0.08$               &  $183\pm14$     &  $102\pm4$\tablenotemark{a}       &    3.8       \\
                       &  $0.43\pm0.08$               &  $297\pm13$     &  $102\pm4$\tablenotemark{a}       &    3.9       \\
H$^{13}$CO$^+\,1-0$    &  $1.26\pm0.09$               &  $176\pm6$      &  $102\pm4$\tablenotemark{a}       &   11.6       \\
                       &  $1.36\pm0.10$               &  $285\pm5$      &  $102\pm4$\tablenotemark{a}       &   12.5       \\
SiO $2-1$              &  $1.52\pm0.11$               &  $182\pm5$      &  $102\pm4$\tablenotemark{a}       &   13.9       \\
                       &  $1.59\pm0.11$               &  $292\pm5$      &  $102\pm4$\tablenotemark{a}       &   14.6       \\
HC$_3$N $J=5-4$        &  $1.05\pm0.15$               &  $191\pm7$      &  $95\pm18$                        &   10.5       \\
CO$^+$ $5/2-3/2 F=2-1$ &  $0.30\pm0.10$               &  191\tablenotemark{c}  &  95\tablenotemark{c}       &   3.0        \\
CO$^+$ $3/2-1/2 F=2-1$ &  $0.17\pm0.10$               &  191\tablenotemark{c}  &  95\tablenotemark{c}       &   1.6        \\
$\rm^{13}CH_3OH$ $5_{0,5}-4_{0.4}$   &  $0.10\pm0.07$ &  191\tablenotemark{c}  &  95\tablenotemark{c}       &   1.0        \\
$\rm^{13}CH_3OH$ $5_{-1,5}-4_{-1.4}$ &  $0.16\pm0.07$ &  191\tablenotemark{c}  &  95\tablenotemark{c}       &   1.6        \\
% $\rm^{13}CH_3OH$ $5_{0,5}-4_{0.4}$   &  $0.24\pm0.07$ &  191\tablenotemark{c}  &  95\tablenotemark{c}       &   2.4        \\
\tableline
\end{tabular}
%\tablecomments{}
\tablenotetext{a}{~Linewidths forced to have the same value in the Gaussian fit.}
\tablenotetext{b}{~$3\sigma$ upper limit assuming a 200 \kms\, linewidth.}
\tablenotetext{c}{~Parameters forced to equal those derived from HC$_3$N Gaussian fit.}
\end{center}
\end{table}

\clearpage

\begin{table}
\begin{center}
\caption{Derived fractional abundances and $\rm H^{13}CO^+$ ratios for each velocity component\label{tab:abunRatios}}
\begin{tabular}{l c c c c}
\tableline
\tableline
%Molecule             &  $X/\rm H_2$      &   ${\rm HCO^+}/X$\,\tablenotemark{a} \\
%\tableline
Molecule             &  $N$                               &      $[X]/[\rm H_2]$\,\tablenotemark{a}  &   ${\rm H^{13}CO^+}/X$ \\
                     &  $(\times10^{13}\,\rm cm^{-2})$    &      $(10^{-10})$                        &                     \\
\tableline
%$\rm H^{13}CO^+ $    & $(6.8\pm1.5)\times10^{-10} \,-\,  (3.8\pm0.5)\times10^{-10}$   &    40               \\
%$\rm HOC^+ $         &                       & $(3.3\pm1.2)\times10^{-10} \,-\,  (2.1\pm0.7)\times10^{-10}$   &    $100\pm30  \,-\, 88\pm19$     \\
%$\rm HCO   $         &                       & $(3.8\pm1.2)\times10^{-9}  \,-\,  (2.0\pm0.5)\times10^{-9} $   &    $7.2\pm1.7 \,-\, 7.5\pm1.7$   \\
$\rm H^{13}CO^+ $    &  $1.6\pm0.5  $        & $4.8\pm1.0 $   &    1               \\
                     &  $1.7\pm0.6  $        & $2.7\pm0.4 $   &    1               \\
$\rm HOC^+ $         &  $0.8\pm0.3  $        & $2.4\pm0.7 $   &    $2.0\pm0.8$   \\
                     &  $1.1\pm0.4  $        & $1.7\pm0.4 $   &    $1.6\pm0.4$   \\
$\rm HCO   $         &  $12.3\pm5.8 $        & $37\pm10 $     &    $0.13\pm0.04$ \\
                     &  $12.9\pm6.1 $        & $20\pm4  $     &    $0.14\pm0.03$ \\
$\rm CO^+$           &  $1.7\pm0.8  $        & $5\pm2   $     &    $0.9\pm0.4$   \\
\tableline
\end{tabular}
\tablenotetext{a}{With $\rm N(H_2)=3.3\pm1.2\times10^{22}\,cm^{-2}$ and $6.3\pm2.1\times10^{22}\rm\,cm^{-2}$ for each velocity component,
respectively, as derived from C$^{18}$O with $\rm^{16}O/^{18}O=150$ \citep{Harrison99} and a CO/H$_2=10^{-4}$.}
%N(C18O)
%2.2\pm0.8\times10^{16}\,cm^{-2}
%4.2\pm1.4\times10^{16}\rm\,cm^{-2}$
% Los errores en la X/H2 solo consideran los errores en el ajuste (10-20%) no los debidos a la incertidumbre en la temp.
%\tablenotetext{b}{Assumed isotopic ratio $\rm^{12}C/^{13}C=40$ \citep{Henkel93}.}
\end{center}
\end{table}

\clearpage

\begin{table}
\begin{center}
\caption{Abundance ratios of $\rm HCO^+$ vs $\rm HOC^+$ and HCO. \label{tab:ratios}}
\begin{tabular}{l c c c c}
\tableline
\tableline
%                           &  NGC\,253                    &  M\,82                            &  NGC\,1068                        &   NGC\,4945     \\
%\tableline
%$\rm [H^{13}CO^+]/[HOC^+]$ &  $2.3\pm0.6$                 &  $\sim2.6$\,\tablenotemark{a}     & $4.4\pm1.0$\,\tablenotemark{b}    &   ...                    \\
%$\rm [H^{13}CO^+]/[HCO]$   &  $0.18\pm0.04  $             &  $0.24\pm0.07$\,\tablenotemark{c} & $0.12\pm0.04$\,\tablenotemark{d}  &   $0.09\pm0.02$\,\tablenotemark{e}  \\
Source                     &   $\rm [HCO^+]/[HOC^+]$            &  $\rm [HCO^+]/[HCO]$              & $\rm [HCO^+]/[CO^+]$          \\
\tableline
NGC\,253                   &    $ 80\pm 30$                     &  $ 5.2\pm 1.8$                    &   $ 38\pm 15$                 \\
                           &    $ 63\pm 17$                     &  $ 5.4\pm 1.3$                    &      ...   \\
%M\,82                      &    $1.9\pm0.9$\,\tablenotemark{a}  &  $0.24\pm0.07$\,\tablenotemark{b} &   $2.3\pm1.0$  \\
M\,82                      &    $ 60\pm 28$\,\tablenotemark{a}  &  $ 9.6\pm 2.8$\,\tablenotemark{b} &   $ 32\pm 16$ \tablenotemark{a} \\
NGC\,1068                  &    $128\pm 28$\,\tablenotemark{c}  &  $ 3.2\pm 1.2$\,\tablenotemark{d} &      ...   \\
NGC\,4945                  &    ...                             &  $ 2.4\pm 1.2$\,\tablenotemark{e} &      ...   \\
\tableline
\multicolumn{4}{c}{\it GC prototypical PDRs} \\
Horsehead                  &    $ 75-200$ \tablenotemark{f}     &   1.1 \tablenotemark{g}           &    $>1800$ \tablenotemark{f}   \\
Orion Bar                  &   $<166-270$ \tablenotemark{h,i}   &   2.4 \tablenotemark{j}           &    $<83-140$  \tablenotemark{i,k}    \\
NGC\,7023                  &    $ 50-120$  \tablenotemark{i}    &    31 \tablenotemark{j}           &      14  \tablenotemark{i}    \\
S140                       &     4110 \tablenotemark{k}         &   $3.5->62$ \tablenotemark{j,l}   &     7420  \tablenotemark{k}    \\
NGC\,2023                  &     940 \tablenotemark{k}          &    1.1 \tablenotemark{j}          &      740  \tablenotemark{k}    \\
M17SW                      &   2262  \tablenotemark{h}          &    ...                            &      250 \tablenotemark{m}    \\
%Mon\,R2 \tablenotemark{i}  &  &  & \\
%NGC\,7027 \tablenotemark{j}&  &  & \\
\tableline
\end{tabular}
%\tablecomments{}
\tablenotetext{a}{\,Derived from single dish data \citep{Mauers91,Fuente06}. See Sect.~\ref{sec.rat} for details.}
% Mauers H13CO+ 1-0   1.5(0.5) K km/s 8 mK  126(37) km/s  W=143(44) km/s
% Fuente  5 mK
\tablenotetext{b}{\,Average ratio from the interferometric maps by \citep{Burillo02}}
% PDRs 0.12 0.04
% NoPDR 0.39 0.11
\tablenotetext{c}{\,Average over the whole line profile in the CND position \citep{Usero04}.}
% Red   3.54 0.95
% Blue  2.90 0.50
\tablenotetext{d}{\,Average value over the three positions in the circunnuclear starburst ring with HCO detections \citep{Usero04}.}
% South position 0.102 0.031
% North position 0.092 0.0354
% East position  0.048 0.0129
\tablenotetext{e}{\,From \citet{Wang04}.}
% H13CO+ 0.59
% HCO    0.38
\tablenotetext{f}{\,\citet{Goico09}}
\tablenotetext{g}{\,\citet{Gerin09}}
\tablenotetext{h}{\,\citet{Apponi99}}
\tablenotetext{i}{\,Orion Bar ionization front and PDR-peak in NGC\,7023 \citet{Fuente03}}
\tablenotetext{j}{\,\citet{Schilke01}}
\tablenotetext{k}{\,\citet{Savage04}}
\tablenotetext{l}{\,\citet{Schene88}}
\tablenotetext{m}{\,\citet{Storzer95}}
\end{center}
\end{table}

\clearpage


\begin{thebibliography}{}
\bibitem[Apponi \& Ziurys(1997)]{Apponi97} Apponi, A.~J., \& Ziurys, L.~M.\ 1997, \apj, 481, 800 
\bibitem[Apponi et al.(1999)]{Apponi99} Apponi, A.~J., Pesch, T.~C., \& Ziurys, L.~M.\ 1999, \apjl, 519, L89 
\bibitem[Bayet et al.(2008)]{Bayet08} Bayet, E., Lintott, C., Viti, S., Mart{\'{\i}}n-Pintado, J., Mart{\'{\i}}n, S., Williams, D.~A., \& Rawlings, J.~M.~C.\ 2008, \apjl, 685, L35 
\bibitem[Bayet et al.(2009)]{Bayet09} Bayet, E., Aladro, R., Mart{\'{\i}}n, S., Viti, S., Mart{\'{\i}}n-Pintado, J. 2009, Submitted to ApJ
\bibitem[Bell et al.(2006)]{Bell06} Bell, T.~A., Roueff, E., Viti, S., \& Williams, D.~A.\ 2006, \mnras, 371, 1865 
%\bibitem[Bell et al.(2007)]{Bell07} Bell, T.~A. 2007, PhD Thesis.
\bibitem[Cappi et al.(1999)]{Cappi99} Cappi, M., et al.\ 1999, \aap, 350, 777
\bibitem[Carral et al.(1994)]{Carral94} Carral, P., Hollenbach, D.~J., Lord, S.~D., Colgan, S.~W.~J., Haas, M.~R., Rubin, R.~H., \& Erickson, E.~F.\ 1994, \apj, 423, 223 
\bibitem[Douglas et al.(1996)]{Douglas96} Douglas, J.~N., Bash, F.~N., Bozyan, F.~A., Torrence, G.~W., \& Wolfe, C.\ 1996, \aj, 111, 1945 
\bibitem[Fuente et al.(2003)]{Fuente03} Fuente, A., Rodr\'{\i}guez-Franco, A., Garc\'{\i}a-Burillo, S., Mart{\i}n-Pintado, J., \& Black, J.~H.\ 2003, \aap, 406, 899 
%\bibitem[Fuente et al.(2005)]{Fuente05} Fuente, A., Garc{\'{\i}}a-Burillo, S., Gerin, M., Teyssier, D., Usero, A., Rizzo, J.~R., \& de Vicente, P.\ 2005, \apjl, 619, L155 
\bibitem[Fuente et al.(2006)]{Fuente06} Fuente, A., Garc{\'{\i}}a-Burillo, S., Gerin, M., Rizzo, J.~R., Usero, A., Teyssier, D., Roueff, E., \& Le Bourlot, J.\ 2006, \apjl, 641, L105 
\bibitem[Garc{\'{\i}}a-Burillo et al.(2000)]{Burillo00} Garc{\'{\i}}a-Burillo, S., Mart{\'{\i}}n-Pintado, J., Fuente, A., \& Neri, R.\ 2000, \aap, 355, 499 
\bibitem[Garc{\'{\i}}a-Burillo et al.(2002)]{Burillo02} Garc{\'{\i}}a-Burillo, S., Mart{\'{\i}}n-Pintado, J., Fuente, A., Usero, A., \& Neri, R.\ 2002, \apjl, 575, L55 
\bibitem[Genzel et al.(1998)]{Genzel98} Genzel, R., et al.\ 1998, \apj, 498, 579 
\bibitem[Gerin et al.(2009)]{Gerin09} Gerin, M., Goicoechea, J.~R., Pety, J., \& Hily-Blant, P.\ 2009, \aap, 494, 977 
\bibitem[Goicoechea et al.(2009)]{Goico09} Goicoechea, J.~R., Pety, J., Gerin, M., Hily-Blant, P., \& Le Bourlot, J.\ 2009, \aap, 498, 771 
\bibitem[Harrison et al.(1999)]{Harrison99} Harrison, A., Henkel, C., \& Russell, A.\ 1999, \mnras, 303, 157 
\bibitem[Henkel et al.(1993)]{Henkel93} Henkel, C., Mauersberger, R., Wiklind, T., Huettemeister, S., Lemme, C., \& Millar, T.~J.\ 1993, \aap, 268, L17 
\bibitem[Henkel et al.(1994)]{Henkel94} Henkel, C., Whiteoak, J.~B., \& Mauersberger, R.\ 1994, \aap, 284, 17
\bibitem[H\"uttemeister et al.(1997)]{Hutte97} H\"uttemeister, S., Mauersberger, R., \& Henkel, C.\ 1997, \aap, 326, 59 
\bibitem[Liszt et al.(2004)]{Liszt04} Liszt, H., Lucas, R., \& Black, J.~H.\ 2004, \aap, 428, 117 
\bibitem[Lord et al.(1996)]{Lord96} Lord, S.~D., Hollenbach, D.~J., Haas, M.~R., Rubin, R.~H., Colgan, S.~W.~J., \& Erickson, E.~F.\ 1996, \apj, 465, 703 
\bibitem[Mart{\'{\i}}n et al.(2003)]{Martin03} Mart{\'{\i}}n, S., Mauersberger, R., Mart{\'{\i}}n-Pintado, J., Garc{\'{\i}}a-Burillo, S., \& Henkel, C.\ 2003, \aap, 411, L465 
\bibitem[Mart{\'{\i}}n et al.(2005)]{Martin05} Mart{\'{\i}}n, S., Mart{\'{\i}}n-Pintado, J., Mauersberger, R., Henkel, C., \& Garc{\'{\i}}a-Burillo, S.\ 2005, \apj, 620, 210 
\bibitem[Mart{\'{\i}}n et al.(2006a)]{Martin06a} Mart{\'{\i}}n, S., Mart{\'{\i}}n-Pintado, J., \& Mauersberger, R.\ 2006a, \aap, 450, L13 
\bibitem[Mart{\'{\i}}n et al.(2006b)]{Martin06b} Mart{\'{\i}}n, S., Mauersberger, R., Mart{\'{\i}}n-Pintado, J., Henkel, C., \& Garc{\'{\i}}a-Burillo, S.\ 2006b, \apjs, 164, 450 
\bibitem[Mart{\'{\i}}n et al.(2008)]{Martin08} Mart{\'{\i}}n, S., Requena-Torres, M.~A., Mart{\'{\i}}n-Pintado, J., \& Mauersberger, R.\ 2008, \apj, 678, 245 
\bibitem[Mart{\'{\i}}n et al.(2009)]{Martin09} Mart{\'{\i}}n, S., Mart{\'{\i}}n-Pintado, J., \& Mauersberger, R.\ 2009, \apj, 694, 610 
\bibitem[Mauersberger \& Henkel(1991)]{Mauers91} Mauersberger, R., \& Henkel, C.\ 1991, \aap, 245, 457 
\bibitem[Mauersberger \& Henkel(1993)]{Mauers93} Mauersberger, R., \& Henkel, C.\ 1993, Reviews in Modern Astronomy, 6, 69
\bibitem[Minh et al.(2007)]{Minh07} Minh, Y.~C., Muller, S., Liu, S.-Y., \& Yoon, T.~S.\ 2007, \apjl, 661, L135
\bibitem[Nguyen et al.(1992)]{Nguyen92} Nguyen, Q.-R., Jackson, J.~M., Henkel, C., Truong, B., \& Mauersberger, R.\ 1992, \apj, 399, 521 
\bibitem[Ott et al.(2005)]{Ott05} Ott, J., Weiss, A., Henkel, C., \& Walter, F.\ 2005, \apj, 629, 767 
\bibitem[Requena-Torres et al.(2006)]{Requena06} Requena-Torres, M.~A., Mart{\'{\i}}n-Pintado, J., Rodr{\'{\i}}guez-Franco, A., Mart{\'{\i}}n, S., Rodr{\'{\i}}guez-Fern{\'a}ndez, N.~J., \& de Vicente, P.\ 2006, \aap, 455, 971 
%\bibitem[Rizzo et al.(2003)]{Rizzo03} Rizzo, J.~R., Fuente, A., Rodr{\'{\i}}guez-Franco, A., \& Garc{\'{\i}}a-Burillo, S.\ 2003, \apjl, 597, L153 
\bibitem[Savage \& Ziurys(2004)]{Savage04} Savage, C., \& Ziurys, L.~M.\ 2004, \apj, 616, 966 
\bibitem[Sage \& Ziurys(1995)]{Sage95} Sage, L.~J., \& Ziurys, L.~M.\ 1995, \apj, 447, 625 
\bibitem[Schenewerk et al.(1988)]{Schene88} Schenewerk, M.~S., Jewell, P.~R., Snyder, L.~E., Hollis, J.~M., \& Ziurys, L.~M.\ 1988, \apj, 328, 785 
\bibitem[Schilke et al.(2001)]{Schilke01} Schilke, P., Pineau des For{\^e}ts, G., Walmsley, C.~M., \& Mart{\'{\i}}n-Pintado, J.\ 2001, \aap, 372, 291 
\bibitem[Spaans \& Meijerink(2007)]{Spaans07} Spaans, M., \& Meijerink, R.\ 2007, \apjl, 664, L23 
\bibitem[Sternberg \& Dalgarno(1995)]{Sternberg95} Sternberg, A., \& Dalgarno, A.\ 1995, \apjs, 99, 565 
\bibitem[St\"orzer et al.(1995)]{Storzer95} St\"orzer, H., Stutzki, J., \& Sternberg, A.\ 1995, \aap, 296, L9 
\bibitem[Usero et al.(2004)]{Usero04} Usero, A., Garc{\'{\i}}a-Burillo, S., Fuente, A., Mart{\'{\i}}n-Pintado, J., \& Rodr{\'{\i}}guez-Fern{\'a}ndez, N.~J.\ 2004, \aap, 419, 897 
\bibitem[van der Tak et al.(2008)]{VdTak08} van der Tak, F.~F.~S., Aalto, S., \& Meijerink, R.\ 2008, \aap, 477, L5 
\bibitem[Wang et al.(2004)]{Wang04} Wang, M., Henkel, C., Chin, Y.-N., Whiteoak, J.~B., Hunt Cunningham, M., Mauersberger, R., \& Muders, D.\ 2004, \aap, 422, 883
\bibitem[Ziurys \& Apponi(1995)]{Ziurys95} Ziurys, L.~M., \& Apponi, A.~J.\ 1995, \apjl, 455, L73 
\end{thebibliography}
\end{document}